\documentclass[conference]{IEEEtran}
\IEEEoverridecommandlockouts
\usepackage{cite}
\usepackage{amsmath,amssymb,amsfonts}
\usepackage{algorithmic}
\usepackage{graphicx}
\usepackage{textcomp}
\usepackage{xcolor}
\usepackage{diagbox}
\usepackage{multirow}

\def\BibTeX{{\rm B\kern-.05em{\sc i\kern-.025em b}\kern-.08em
    T\kern-.1667em\lower.7ex\hbox{E}\kern-.125emX}}
\begin{document}

\title{Task-aware Warping Factors in Mask-based Speech Enhancement}

\author{\IEEEauthorblockN{Qiongqiong Wang, Kong Aik Lee\textsuperscript{*}, Takafumi Koshinaka\textsuperscript{*}, Koji Okabe, and Hitoshi Yamamoto\thanks{*Kong Aik Lee and Takafumi Koshinaka are currently with the Institute for Infocomm Research, A*STAR, Singapore and the School of Data Science, Yokohama City University, Japan, respectively.}}
\IEEEauthorblockA{\textit{Biometrics Research Laboratories, NEC Corporation}, 
Japan \\
q-wang@nec.com}

}

\maketitle

\begin{abstract}
This paper proposes the use of two task-aware warping factors in mask-based speech enhancement (SE). 
One controls the balance between speech-maintenance and noise-removal in training phases,
while the other controls the degree of enhancement applied to specific downstream tasks in testing phases. 
Our proposal is based on the observation that SE systems trained to improve speech quality 
often fail to improve other downstream tasks, 
such as automatic speaker verification (ASV) and automatic speech recognition (ASR), because they do not share the same objectives. 
It is easy to apply the proposed dual-warping factors approach to any mask-based SE method,
and it allows a single SE base module to handle multiple tasks without task-dependent training. 
The effectiveness of our proposed approach has been confirmed on 
the SITW dataset for ASV evaluation 
and the LibriSpeech test-clean set for ASR and speech quality evaluations
of 0-20dB. 
We show that different warping values are necessary in the testing phases
for a single SE base module to achieve optimal performance w.r.t. the three tasks.
With the use of task-aware warping factors, speech quality 
was improved 
by an $84.7\%$ PESQ increase, while ASV 
had a $22.4\%$ EER reduction, 
and ASR 
had a $52.2\%$ WER reduction, on 0dB speech.   
The effectiveness of the task-aware warping factors were also cross-validated on VoxCeleb-1 test set for ASV and LibriSpeech dev-clean set for ASR and quality evaluations.
The proposed method is highly effective and easy to apply in practice. 
\end{abstract}


\section{Introduction}
Single-channel speech enhancement (SE) aims at improving the quality
and intelligibility of speech signals degraded by additive noise,
in order to improve human or machine perception of speech\cite{loizou2013}.
It has attracted much attention due to its importance in real-world applications,
including hearing aids\cite{Park2020}, mobile communication, automatic speech recognition (ASR)\cite{MOORE2017574}, and automatic speaker verification (ASV)\cite{Shi2020}.

Numerous SE methods have been proposed over the past decades. 
They can be classified into two categories, in accord with the signal processing domain on which they work. 
Time-domain methods operate directly on one-dimensional raw waveforms
of speech signals \cite{Pascual2017}\cite{Rethage2018},
while frequency-domain methods manipulate two-dimensional speech spectrograms 
\cite{Wang2014,Deliang2017,Xu2014-regression,Tan2019,Kishore2020,Yin2020}.
In the latter, two types of training targets are commonly used\cite{Deliang2017}:  
(i) mapping-based targets corresponding to the spectral representations of clean speech \cite{Xu2014-regression}\cite{Tan2019}
and 
(ii) masking-based targets, i.e., the main stream method
of predicting a time-frequency (T-F) mask over input spectrograms \cite{ Kishore2020}\cite{Yin2020}.
Mask-based approaches, generally perform significantly better than mapping-based ones for supervised speech enhancement \cite{Deliang2017}.

The T-F masks used in SE could be real-value, e.g., 
ideal ratio masks (IRMs)\cite{Narayanan2013} and spectral magnitude masks (SMMs)\cite{Wang2014}.
Complex-value masks have also been proposed to take phase information into consideration, 
e.g., phase-sensitive masks (PSMs) \cite{Erdogan2015} 
and complex ideal ratio masks (cIRMs)\cite{Williamson2016},
which have been the subject of increased interest because of their potential for use in quality-oriented SE.
Prior work in ASV and ASR have shown that phase information is not necessarily useful but redundant in some cases in recognition tasks \cite{Dengli}, and,  for that reason, this paper focuses on real-value masks.

It has been shown that SE optimized for improvement of human perception often fails to improve performance 
in machine-oriented downstream tasks, such as ASV and ASR \cite{Shi2020}\cite{Shon2019}\cite{Sadjadi2010}.
That is because the objective of SE is to improve speech quality by suppressing noise and 
offers no guarantee w.r.t. downstream tasks. 
In fact, the artifacts and distortions induced by SE might even
deteriorate their performance.
The study in \cite{Shon2019} proposed VoiceID loss based on the feedback from a speaker verification
model in SE training, in order to improve robustness of ASV. 
However, the enhanced speech showed consistently worse speech quality than that with conventional methods \cite{Novotny2018-1}. 
It was pointed out that better speech quality
does not necessarily imply better speaker verification.


This observation motivated us to investigate approaches to the training of a SE net for different downstream tasks, including ASV, speech quality, and ASR.
We also seek a new SE approach that takes into account downstream task performance and computational costs.
Specifically, our contributions in this paper are as follows. 
Firstly, we propose the use of dual-warping factors for optimal performance without the need of task-dependent training.
Secondly, we show that the best performance for ASV, ASR, and perceptual quality can be achieved 
using different warping factors for a single SE base module, and the improvements are significant.

The remainder of this paper is organized as follows. 
Section 2 introduces T-F masking methods.
Section 3 presents the proposed warping factors, as well as 
the neural network structure used.
Section 4 describes our experimental setup, results, and analyses.
Section 5 summarizes our work. 

\section{T-F Domain Masking Methods}
\label{sec:mask}
\subsection{Maintaining the Integrity of the Specifications}

Time-frequency (T-F) masking applies a two-dimensional mask to the T-F representation of a source mixture in order to extract the target source.
The most widely used T-F representation is computed using the short-time
Fourier transform (STFT), and can be converted backward to a time-domain signal with the inverse STFT.
There are two categories of T-F masks:
real-value masks and complex-value masks.
In this paper, we advocate the use of real-valued masks, specifically,  
the ideal ratio masks (IRMs)\cite{Narayanan2013}:
\begin{equation}
\label{eq:irm}
M_{\text{irm}}(t,f)=\left(\frac{S^2(t,f)}{S^2(t,f)+N^2(t,f)}\right)^\beta,
\end{equation}
where $S^2(t,f)$ and $N^2(t,f)$ denote energy of speech and noise at time-frequency bin $(t,f)$, respectively. 
The tunable parameter $\beta$ scales the mask, and is commonly chosen as $0.5$. %
Mean square error (MSE) is typically used as the cost function for IRM estimation \cite{Deliang2017}. 
The enhanced amplitude spectrogram is combined with
the noisy phase spectrogram to produce enhanced speech.

\section{Proposed method}
\begin{figure}[t]
\includegraphics[width=\columnwidth]{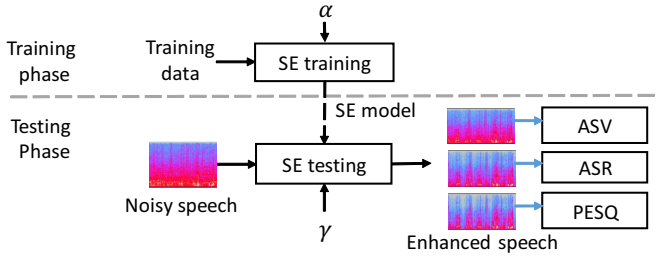}
\caption{Scenario of proposed SE application to downstream tasks; $\gamma$ is an application-dependent testing warping factor; $\alpha$ is an application-independent training warping factor. Here, the speech perceptual quality task is referred to as ``PESQ'' a.k.a. the Perceptual Evaluation of Speech Quality.
}
\label{fig:setask}
  \vspace*{-2mm}
\end{figure}

In this section, we present the task-aware dual-warping factors, which are used to enable a single SE base module to work on multiple tasks. 
We investigate three tasks: speech perceptual quality, ASV, and ASR, 
as shown in Fig.~\ref{fig:setask}.

\subsection{Modification of IRM}\label{ssec:newirm}
Two warping factors are employed on the masks. 
One is a training warping factor $\alpha$ that adjusts the balance between speech-maintenance 
and noise removal in the learning of neural networks.
With the training warping factor $\alpha$, the learning target in a SE neural network is 
\begin{equation}
\label{eq:mask_train}
M_{\text{tr}}(t,f):=\left(\frac{S^2(t,f)}{S^2(t,f)+N^2(t,f)}\right) ^\alpha.
\end{equation}
The relationship with IRM definition in \eqref{eq:irm} is 
$M_{\text{tr}}=M_{\text{irm}}^\frac{\alpha}{\beta}$.
$\alpha$ affects model learning.
A larger $\alpha$ gives greater degree of emphasis to learning higher mask values, i.e., mask elements in which the speech to noise ratio (SNR) is higher.
That is, it gives priority to speech maintenance.
By way of contrast, a smaller $\alpha$ gives lighter degree of emphasis to learning higher mask values.
Rather, it gives greater degree of emphasis to learning mask elements of smaller values.
That is, it gives priority to noise removal.

We also propose a testing warping factor $\gamma$ that controls the degree
of enhancement in the testing phases.
It is defined as 
\begin{equation}
\label{eq:mask_eval_def}
M_{\text{te}}(t,f):=\left(\frac{S^2(t,f)}{S^2(t,f)+N^2(t,f)}\right)^\gamma.
\end{equation}
In the testing phase where there are no ground truths, the mask value $\hat{M}_{\text{te}}(t,f)$ is calculated by applying exponent $\frac{\gamma}{\alpha}$ to $\hat{M}_{\text{tr}}(t,f)$ which is produced by the trained network: 
\begin{equation}
\label{eq:mask_eval}
\hat{M}_{\text{te}}(t,f)=\hat{M}^{\frac{\gamma}{\alpha}}_{\text{tr}}(t,f),
\end{equation}
The log-power spectra (LPS) of the enhanced speech is obtained as
\begin{equation}
\log|S(t,f)|=
\log|Y(t,f)|+
\log|\hat{M}_{\text{te}}(t,f)|,
\end{equation}
where $Y(t,f)$ denotes the noisy speech magnitude. 
It can be tuned for specific downstream tasks.
A smaller $\gamma$ denotes the lower degree
of enhancement, which maintains more original signals.
When $\gamma$ is $0$, it falls back to the case in which no SE is applied.
When we use the two warping factors together, $\beta$ in the IRM definition in \eqref{eq:irm} becomes implicit. 

\subsection{Densely connected BLSTM (D-BLSTM)}
\label{ssec:densenet}
\begin{figure}[t]
\centering
\includegraphics[width=0.85\columnwidth]{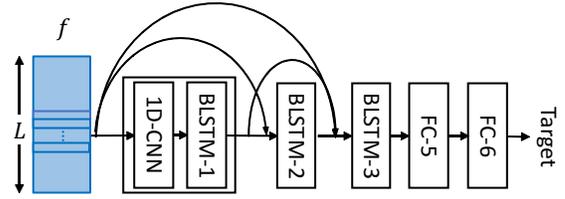}
\caption{{ Densely connected BLSTM (D-BLSTM) structure}}
\label{fig:dnn}
 \vspace*{-2mm}
\end{figure}

In recent years, deep learning techniques
in SE have become increasingly popular \cite{Xu2014,Valin2018,Xia2020}.
To model time sequences, recurrent neural networks (RNN)
have an inherent advantage due to their use of recursive structures between
previous and current frames in order to obtain long-term
contextual information \cite{Valin2018,Xia2020,Sun2018}. 
%
DenseNets have shown a number of compelling advantages w.r.t. strengthening feature propagation and encouraging feature reuse \cite{Huang2019}.
We use a densely connected Bidirectional LSTM structure here to illustrate the potential of the proposed warping factors; see Fig.~\ref{fig:dnn}.

The network is trained to predict $M_\text{tr}$ in \eqref{eq:mask_train},
given the input noisy LPS features.
A 1-D convolutional layer with kernel size of $2n+1$ collects context information for $[t-n, t+n]$ frames
and gives outputs of $f$ dimension to input to the first BLSTM layer,
where $f$ represents acoustic feature dimension,
and $n$ determines the temporal context. 
The output dimension of the three following BLSTM layers remains the same as that of $f$.
Due to the dense connection between the input and the three BLSTM layers, 
the inputs of BLSTM-2 and BLSTM-3 have dimensionality of $2f$ and $3f$, respectively. 
The output of BLSTM-3 is fed into two fully-connected layers with $f$ nodes. 
In order to train a robust SE model, noisy speech of multiple SNR values is commonly used.

\section{Experiments}
\label{sec:exp}
We investigated the effect of the proposed warping factors in SE w.r.t. noisy speech of 
0dB, 10dB, and 20dB,
for three downstream tasks: ASV, ASR, and speech quality evaluation.
The datasets used in the experiments are summarized in  Table~\ref{tab:setting}.

\subsection{Experimental settings}
\label{ssec:setting}
\subsubsection{SE experimental settings}
\label{sssec:setting1}
We utilized the Deep Noise Suppression (DNS) Challenge \cite{dns} for SE training data.
The clean speech dataset is selected from the public audio books dataset LibriVox \cite{librivox}.
It contains about 500 hours of speech from 2,150 speakers.
All the  clips are 31 seconds long.
We used the AudioSet \cite{Gemmeke2017} noise clips in the DNS Challenge,
which contains about 150 audio classes and 60,000 clips.
Noisy clips of 
0dB, 5dB, and 10dB
were generated by combining individual clips from the clean set with a clip randomly chosen from the noise set.

In training, 257-dimension LPS features of randomly chosen 8-sec segments from noisy clips were fed into the network.
The 1-D convolutional layer had the kernel size of $7$ and output dimension is 257.
Thus, the input includes context of 7 neighbouring frames $(\pm 3)$.
All the three BLSTM layers' output dimensions were 257 and input dimensions were $257$, $257\times2$ and $257\times3$, respectively, due to the dense connection.
The number of memory cells in each BLSTM layers was 512.
Each of the two fully connected layers had 257 nodes.
We used an Adam optimizer.
Minibatch size was 80.
The learning rate was initially set as 0.001 and reduced by $20\%$ after each epoch. 
MSE was used as the loss function between $M_\text{tr}$ in \eqref{eq:mask_train} and outputs. 
All models were trained with 15 epochs. 

\subsubsection{Downstream tasks settings}
\label{sssec:setting2}

\begin{table}[t]
\caption{Datasets used in the experiments; SITW* represents SITW core-core \emph{eval} set }
\begin{center}
\begin{tabular}{p{1.3 cm}| p{1.3 cm} |  p{1.2 cm}| p{1.3 cm} |p{1.3cm}}
\hline
Downstream task        & SE train & Train & 1st test  & 2nd test \\
 \hline
ASV &\multirow{3}[1]{1.3cm}{LivbriVox +AudioSet  (DNS Challenge)} &
VoxCeleb-1,2 train & SITW* + MUSAN &VoxCeleb-1 test + PRISM  \\ \cline{3-5}\cline{1-1}
ASR&& \multirow{2}[1]{1.3cm}{Pre-trained \cite{watanabe2018espnet}} & \multirow{2}[1]{1.3cm}{LibriSpeech test-clean + MUSAN} & \multirow{2}[1]{1.3cm}{LibriSpeech dev-clean + PRISM} \\\cline{1-1}
Speech quality&& &  & \\\cline{3-3}
\hline
\end{tabular}
\label{tab:setting}
\end{center}
\vspace*{-3mm}
\end{table}

We prepared two testing sets for each downstream task,
i.e., the first one for finding the optimal warping-factor setting and the second one for validating the setting,
to demonstrate the effectiveness of the proposed method. 
ASV evaluations were conducted on the Speakers in the Wild (SITW) database \cite{sitw}, core-core, \emph{eval} set and VoxCeleb-1 \emph{test} set.
We used the \emph{train} set of VoxCeleb 1 and 2 corpora \cite{vox} to train a standard x-vector \cite{Snyder2018} extractor. 
We evaluated ASR on the LibriSpeech \emph{train-clean} and \emph{dev-clean} datasets with a pre-trained Transformer model for ESPnet,
an end-to-end speech processing toolkit \cite{watanabe2018espnet}. 
The same LibriSpeech datasets were used in speech quality evaluation.
For all three tasks, we created noisy speech of 
0dB, 10dB, and 20dB 
by combining their respective test sets with noise. 
For the 1$st$ test set, we used the noise category of the MUSAN dataset \cite{musan}.
For the 2$nd$ test set, we used noise samples in the PRISM corpus \cite{Ferrer2011}.
Performance w.r.t. ASV, ASR, and speech quality was evaluated using, as measures, equal error rate (EER), word error rate (WER) and Perceptual Evaluation of Speech Quality (PESQ) \cite{pesq}, respectively. 

\subsection{Results and analysis}
\label{sec:results}
\begin{table}[t]
\caption{PESQ scores and ASV EER($\%$) using conventional SE methods and D-BLSTM-base structure }
\begin{center}
\begin{tabular}{l| c c c |c c c}
\hline
                            &\multicolumn{3}{c|}{PESQ} & \multicolumn{3}{c}{ASV EER $(\%)$ }\\
                           & 0dB   & 10dB  & 20dB      & 0dB   & 10dB  & 20dB    \\
 \hline
No enh                      & 1.24  & 1.65  & 2.50     & 5.85  & \textbf{4.43}  & \textbf{4.16}  \\
RNNoise \cite{Valin2018}    & 1.30  & 1.76  & 2.54     & 6.26  & 4.68  & 4.29  \\
NSNet \cite{dns}            & 1.34  & 1.83  & 2.56     & 11.18  & 8.56  & 7.52  \\
D-BLSTM-base                & \textbf{2.18}  & \textbf{2.89}  & \textbf{3.46}     & \textbf{5.28}  & 4.54  & 4.48  \\
\hline
\end{tabular}
\label{tab:tab1}
\end{center}
\end{table}
\begin{table}[t]
\vspace*{-5mm}
\caption{ASV performance with varying $\alpha$ and a fixed $\gamma=0.5$}
\begin{center}
    \begin{tabular}{l| c c c c c}
\hline
$\alpha$        & 0.25  & 0.50  & 1.00      & 1.50   & 2.00    \\
 \hline
20dB            & 4.51  & 4.48  & 4.07     & \textbf{3.96}  & 3.99 \\
10dB            & 4.84  & 4.54  & 4.18     & \textbf{4.13}  & 4.24  \\
0dB             &   -   & 5.28  & \textbf{4.78}     & \textbf{4.78}  & 4.87  \\
\hline
\end{tabular}
\label{tab:alpha}
\end{center}
\vspace*{-5mm}
\end{table}

\begin{table*}[!t]
\caption{ASV EER($\%$) with varying testing warping factors $\gamma_v$ and $\gamma_f$ in SE for VAD and feature extraction, respectively }
\hskip -0.5cm
\fontsize{8.5}{10}\selectfont 
\mbox{}\hfill
\begin{minipage}[t]{.35\linewidth}
\centering
\begin{tabular}{p{0.08\linewidth}|p{0.05\linewidth}p{0.05\linewidth}p{0.05\linewidth}p{0.05\linewidth}p{0.05\linewidth}p{0.05\linewidth}}
\multicolumn{7}{c}{0dB}\\
 \hline 
  \diagbox[width=3.0em,height=1.5em,innerleftsep=1.5pt,innerrightsep=1pt]{$\gamma_v$}{$\gamma_f$}    
      &  0   &  0.375  &  0.50    &  0.75     & 1.50       &  3.00 \\
 \hline
 0 &  5.85 &  4.95 &  4.81         &  4.87      &  5.77     &  7.63   \\ 
 0.375 &  5.85 &  4.92 &  4.78         &  4.84      &  5.71     &  7.60   \\
 0.50 &  5.82 &  4.89 &  4.78         &  4.84      &  5.58     &  7.46   \\
 0.75 &  5.69 &  4.81 &  4.68         &  4.76      &  5.67     &  7.19   \\
 1.50 &  5.55 &  4.68 & 4.59 &  \textbf{4.54}      &  5.17     &  7.27  \\
 3.00 &   -   &  -    & -             & -          &  -        &  -   \\
 \hline
\end{tabular}
\end{minipage}\hfill
 \hspace{-0.02\linewidth}
\begin{minipage}[t]{.35\linewidth}
\centering
\begin{tabular}{p{0.08\linewidth}|p{0.05\linewidth}p{0.05\linewidth}p{0.05\linewidth}p{0.05\linewidth}p{0.05\linewidth}p{0.05\linewidth}p{0.05\linewidth}p{0.05\linewidth}}
\multicolumn{7}{c}{10dB}\\ \hline 
 \diagbox[width=3.0em,height=1.5em,innerleftsep=1.5pt,innerrightsep=1pt]{$\gamma_v$}{$\gamma_f$}
       &  0   &  0.375  &  0.50    &  0.75     & 1.50       &  3.00 \\
 \hline
 0  &  4.43   &  4.24   &  4.13   &  4.24      &  4.89     &  6.23   \\
 0.375  &  4.46   &  4.24   &  4.16   &  4.21      &  4.84     &  6.12   \\
 0.50  &  4.40   &  4.18   &  4.13   &  4.18      &  4.76     &  5.99   \\
 0.75  &  4.43   &  4.05   &  4.05   &  4.10      &  4.65     &  5.85   \\
 1.50  &  4.35   &  4.02   &  \textbf{3.96}   &  \textbf{3.96}    &  4.48     &  5.91   \\
 3.00  &  4.37   &  4.05   &  4.03   &  \textbf{3.96}      &  4.37     &  5.93   \\
 \hline
\end{tabular}
  \end{minipage}\hfill
\hspace{-0.02\linewidth}
\begin{minipage}[t]{.35\linewidth}
\centering
\begin{tabular}{p{0.08\linewidth}|p{0.05\linewidth}p{0.05\linewidth}p{0.05\linewidth}p{0.05\linewidth}p{0.05\linewidth}p{0.05\linewidth}p{0.05\linewidth}p{0.05\linewidth}}
\multicolumn{7}{c}{20dB}\\
\hline 
\diagbox[width=3.0em,height=1.5em,innerleftsep=1.5pt,innerrightsep=1pt]{$\gamma_v$}{$\gamma_f$}           &  0   &  0.375  &  0.50    &  0.75     & 1.50       &  3.00 \\
 \hline
 0  &  4.16   &  4.05   &  4.05   &  4.18      &  4.68     &  5.91   \\
 0.375  &  4.13   &  3.96   &  4.02   &  4.16      &  4.59     &  5.77   \\
 0.50  &  4.10   &  3.91   &  3.96   &  4.05      &  4.54     &  5.69   \\
 0.75  &  4.16   &  3.86   &  3.88   &  3.99      &  4.48     &  5.58   \\
 1.50  &  4.05   &  3.88   &  3.88   &  4.02      &  4.43     &  5.44   \\
 3.00  &  4.02   &  \textbf{3.83}   &  \textbf{3.83}   &  3.96      &  4.32     &  5.47   \\
 \hline
\end{tabular}

  \end{minipage}\hfill
  
  \label{tab:vadfeat}
\mbox{}
\vspace*{-3mm}
\end{table*}

We first investigated performance of the conventional methods RNNoise \cite{Valin2018} and NSNet \cite{dns}, 
as well as the use of a D-BLSTM-base structure (the model shown in Fig.~\ref{fig:dnn}) without warping factors, on PESQ and ASV,
as shown in Table~\ref{tab:tab1}.
The experiments were conducted on the first testing sets.
All three SE methods consistently improved the PESQ of all three SNR sets, 
and use of the D-BLSTM-base structure was particularly effective,
though it resulted in reduced ASV EER only on the 0dB set to a small extent.
This further illustrates the problem that SE developed to improve speech quality often fails to work well in machine-oriented tasks.

We next evaluated the effects of the proposed two warping factors independently on the same ASV datasets. 
Table~\ref{tab:alpha} shows EER of ASV 
following SE with varying training warping factor $\alpha$ and a fixed testing warping factor $\gamma=0.5$.
The result for 0dB set when $\alpha=0.25$ was not obtainable because VAD removed too many frames after SE.
The setting of $\alpha=\gamma=0.5$ is the baseline used in conventional methods \cite{Deliang2017} and also D-BLSTM-base in Table~\ref{tab:tab1}.
Unlike the results shown in Table~\ref{tab:tab1}, $\alpha$ values other than $0.5$ successfully improved ASV performance.
The best performance was 
achieved using $\alpha=1.5$,
which reduced EER by $18.3\%$, $6.8\%$, $4.8\%$ in ASV for, respectively,
0dB, 10dB, and 20dB sets,
as compared with those without SE.
A further increase of $\alpha$ ($\alpha=2$), however, degraded ASV performance.
Since the training warping factor controls the balance between speech maintenance and noise removal in training, 
we concluded that a good balance in SE for ASV could be achieved by giving greater priority 
to speech maintenance rather than noise removal.
\begin{figure}[!t]
\vspace*{-1mm}
\centering
\includegraphics[width=0.7\linewidth]{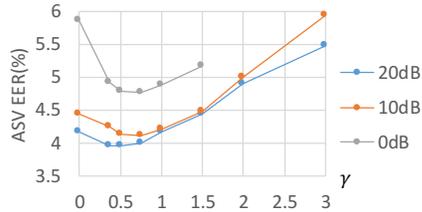}
     \vspace*{-3mm}
     \caption{ASV performance following SE using varying $\gamma$ and a fixed $\alpha=1.5$}
    \label{fig:result_asv}
   
\end{figure}
\begin{figure}[!t]
\vspace*{-5mm}
\centering
\includegraphics[width=0.7\linewidth]{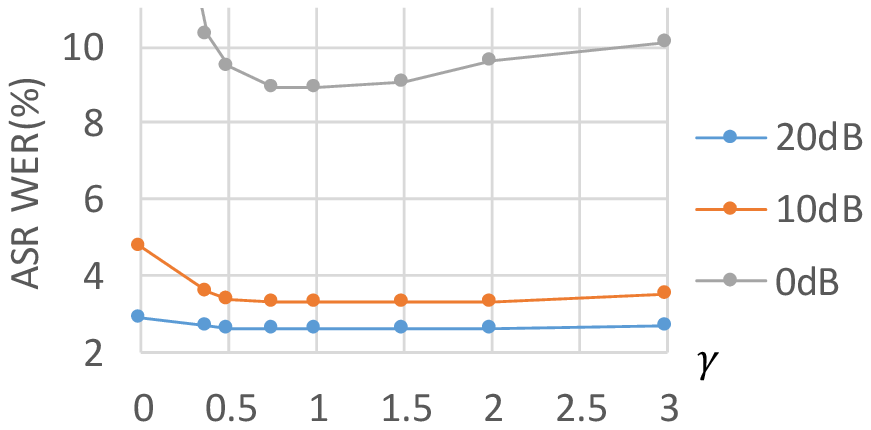}
        \vspace*{-3mm}
\caption{ASR performance following SE using varying $\gamma$ and a fixed $\alpha=1.5$}
    \label{fig:result_asr}
\end{figure}
\begin{figure}[!t]
\vspace*{-5mm}
\centering
\includegraphics[width=0.7\linewidth]{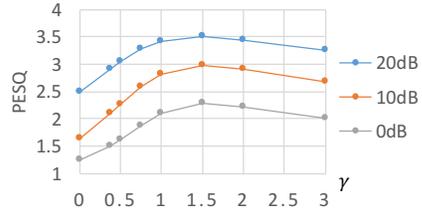}
        \vspace*{-3mm}
    \caption{Speech quality (PESQ) following SE using varying $\gamma$ and a fixed $\alpha=1.5$}
    \label{fig:result_pesq}
    \vspace*{-3mm}
\end{figure}

\begin{figure}[!t]
\centering
\includegraphics[width=0.8\linewidth]{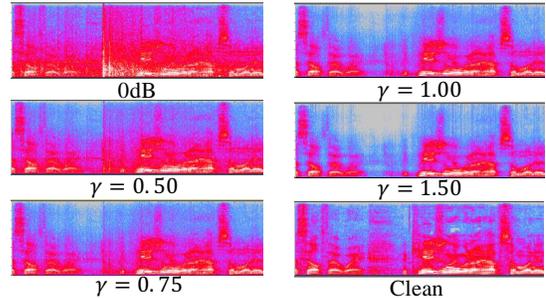}
    \caption{Illustration of spectrogram examples of a 2.4-sec 0dB segment from SITW core-core \emph{eval} set with additional MUSAN noise, that after speech enhancement and the clean one; $\alpha$ in speech enhancement is set $1.5$.}
    \label{fig:spec}
\vspace*{-5mm}
\end{figure}

We next fixed the training warping factor at $\alpha=1.5$ and investigated SE with varying testing warping factors for ASV; see Fig~\ref{fig:result_asv}.
SE using a smaller testing warping factor benefits ASV. 
EER was reduced by $18.6\%, 7.4\%$ and $4.1\%$ with $\gamma=0.75$ in ASV for, respectively, 
0dB, 10dB, and 20dB sets,
as compared with those without SE.
Since the testing warping factor controls the degree of enhancement applied to ASV, 
a larger $\gamma$ may introduce too much distortion at the same time when noise is well removed.
We concluded that a weaker SE benefited ASV more, and that distortion was especially harmful to ASV.
In addition, we observed that a shift to a smaller $\gamma$ offered the best performance when the speech became noisier.
%
Illustration of speech spectrogram examples is shown in Fig.~\ref{fig:spec}.
Clearer denoising effects were observed with increases of the testing warping factor $\gamma$.
We subsequently, used a single SE model ($\alpha=1.5$) and focused on the testing warping factor for easier implementation in practice.

Table~\ref{tab:vadfeat} shows the effect of the testing warping factor for two sub-tasks in ASV: VAD and acoustic feature extraction. 
$\gamma_v=0$ and $\gamma_f=0$ stand for no SE applied for, respectively, VAD and feature extraction. 
Experimental results show that the use of a larger testing warping factor value enhanced speech from which a better VAD was obtained for ASV, 
while the use of a smaller value enhanced speech from which better ASV features were extracted.
 Feature extraction was more sensitive to SE than that with VAD.
The use of such optimal SE on only VAD and on only features yielded reductions of up to $5.1\%$ and $17.8\%$ EER, respectively, for ASV on 0dB speech.
The combination of VAD and features from enhanced speech, using their respective optimal SEs in a single ASV evaluation, yielded, as expected, a further improvement, specifically, with $22.4\%$ EER reduction. 
By having the same $\alpha$ we were able to save on computation costs during testing phases, as both VAD and feature extraction were based on the same masks.

We further extended the investigation to SE with ASR and speech quality evaluation
 using a fixed training warping factor ($\alpha=1.5$) and varying the testing warping factor $\gamma$.  
 Note that we reused, for simplicity, the value $\alpha=1.5$, which had been determined through the previous ASV experiments and may not be optimal for the other two tasks.  
 Investigation of the dependence of $\alpha$ on downstream tasks is included in our future work.
Fig.~\ref{fig:result_asr} shows ASR performance w.r.t. enhanced and original noisy speech.
For all of the testing warping factor values that we investigated, ASR performance was consistently improved.
In the evaluation of 
0dB, 10dB, and 20dB sets,
WER values were reduced by up to $52.2\%$, $31.3\%$ and $10.3\%$ from $18.6\%$, $4.8\%$ and $2.9\%$ WER without SE, respectively.
ASR was much less sensitive to warping factor values than was ASV.
The best ASR performance was obtained with warping factor $\gamma=0.75\sim1.00$.

Fig.~\ref{fig:result_pesq} shows speech-quality measure PESQ scores for enhanced and original noisy speech.
As in ASR, SE with any investigated warping factor improved speech quality to some degree. 
Its sensitivity to $\gamma$ was between those for ASV and ASR. 
The PESQ scores were increased from $1.24$, $1.65$ and $2.50$ to $2.29$, $2.99$ and $3.50$ for   
0dB, 10dB, and 20dB sets,
respectively, using SE with $\gamma=1.50$.
The optimal $\gamma$ value was the largest among the three tasks. 


\begin{table}[t]
\caption{Cross-validation experiments.
"-": no SE was applied; "SE*": SE with the same network structure and using the testing warping factors as $\gamma_{ASV}=0.75$, $\gamma_{ASR}=1.0$, and $\gamma_{pesq}=1.5$ which were optimized in the $1^{st}$ test set.}
\begin{center}
    \begin{tabular}{p{0.08\linewidth}|p{0.05\linewidth}p{0.05\linewidth}|p{0.05\linewidth}p{0.05\linewidth}|p{0.05\linewidth}p{0.05\linewidth}}
\hline
     SNR          &\multicolumn{2}{c|}{PESQ} & \multicolumn{2}{c|}{ASV EER $(\%)$} & \multicolumn{2}{c}{ASR WER $(\%)$ } \\
                 & -        & SE*               & -   & SE*             & -    & SE*    \\
 \hline
20dB             & 2.04    & \textbf{3.32}     & 3.93  & \textbf{3.92}   & 2.6  &   \textbf{2.4} \\
10dB             & 1.28    & \textbf{2.59}     & 6.89  & \textbf{5.75}   & 6.8  &   4.5\\
0dB              & 1.08    & \textbf{1.69}     & 19.89 & \textbf{12.50}  & 50.3  &  28.1 \\
\hline
\end{tabular}
\label{tab:tab3}
\end{center}
\vspace*{-5mm}
\end{table}

In table~\ref{tab:tab3}, we validated the results on the second test set
using the warping factor setting optimized through the previous experiments using the first test set.
We compared the downstream task performance without speech enhancement and the proposed method. 
The performance were successfully improved at all of 0dB, 10dB, and 20dB SNRs all over the three downstream tasks after we applied the proposed speech enhancement. 
Therefore, it verified the generalization property of the proposed algorithm.

\section{Summary}\label{SCM}
This paper has presented the use of warping factors for mask-based speech enhancement (SE) 
applicable to multiple downstream tasks without task-dependent training.
Evaluations of its effectiveness have been conducted for ASV, ASR for machines, and speech quality w.r.t. human hearing. 
Experiment results showed that different warping factor values used in the
testing phases achieved respective optimal performance levels for different tasks over various SNRs
with significant improvements: speech perceptual quality was improved by a $84.7\%$ PESQ increase;  EERs of ASV were reduced by $22.4\%$; and WERs of ASR were reduced by $52.2\%$, on 0dB speech. In addition, ASV performance was more sensitive, while ASR and speech quality evaluations accepted a wider range of warping factors and showed a significant improvement over all testing warping factor values investigated.
The generalization property of the proposed algorithm is also verified in another datasets.
The proposed method is effective and easy to implement in practice to any mask-based SE system.
For  future work, we will explore the optimization of the warping factors by gradient descent directly.

\end{document}